\documentclass[12pt,preprint]{aastex}
\usepackage{longtable}
\usepackage{txfonts}
\usepackage{lscape}
\usepackage{color}
\newcommand{\suzaku}{{\it Suzaku}}
\newcommand{\ktbb}{$kT_{\rm bb}$}
\newcommand{\ktin}{$kT_{\rm in}$}
\newcommand{\rbb}{$R_{\rm bb}$}
\newcommand{\rin}{$R_{\rm in}$}
\newcommand{\kte}{$kT_{\rm e}$}

\begin{document}

\shorttitle{\suzaku~Observation of EXO 0748$-$676}
\shortauthors{Zhang et al.}

\title {\suzaku~Observation of the High-Inclination Binary EXO 0748$-$676 in the Hard State}

\author{Zhongli Zhang\altaffilmark{1,2},
        Soki Sakurai\altaffilmark{1},
        Kazuo Makishima\altaffilmark{1,3,4},
	Kazuhiro Nakazawa\altaffilmark{1,4},
        Ko Ono\altaffilmark{1}, 
        Shin'ya Yamada\altaffilmark{5},
	and Haiguang Xu\altaffilmark{6,7}
	}

\altaffiltext{1}{Department of Physics, School of Science, The University of Tokyo, 7-3-1, Hongo, Bunkyo-ku, Tokyo 113-0033, Japan; e-mail: zzhang@juno.phys.s.u-tokyo.ac.jp}
\altaffiltext{2}{Shanghai Astronomical Observatory, Chinese Academy of Sciences, 80 Nandan Road, 200030 Shanghai, PR China}
\altaffiltext{3}{MAXI Team, Institute of Physical and Chemical Research (RIKEN), Wako, Saitama 351-0198, Japan}
\altaffiltext{4}{Research Center for the Early Universe, The University of Tokyo, 7-3-1, Hongo, Bunkyo-ku, Tokyo 113-0033, Japan}
\altaffiltext{5}{Department of Physics, Tokyo Metropolitan University, Minami-Osawa 1-1, Hachioji-shi, Tokyo 192-0397, Japan}
\altaffiltext{6}{Department of Physics and Astronomy, Shanghai Jiao Tong University, 800 Dongchuan Road, Minhang, Shanghai 200240, China}
\altaffiltext{7}{IFSA Collaborative Innovation Center, Shanghai Jiao Tong University, 800 Dongchuan Road, Minhang, Shanghai 200240, China}

%%%%%%%%%%%%%%%%
\begin{abstract}
%%%%%%%%%%%%%%%%

Utilizing an archived \suzaku~data acquired on 2007 December 25 for 46 ks, X-ray spectroscopic properties of the dipping and eclipsing low-mass X-ray binary EXO 0748$-$676 were studied. At an assumed distance of 7.1 kpc, the data gave a persistent unabsorbed luminosity of $3.4\times10^{36}$ erg cm$^{-2}$ s$^{-1}$ in 0.6 $-$ 55 keV. The source was in a relatively bright low/hard state, wherein the 0.6 $-$ 55 keV spectrum can be successfully explained by a ``double-seed" Comptonization model, incorporating a common corona with an electron temperature of $\sim13$ keV. The seed photons are thought to be supplied from both the neutron star surface, and a cooler truncated disk. Compared to a sample of non-dipping low-mass X-ray binaries in the low/hard state, the spectrum is subject to stronger Comptonization, with a relatively larger Comptonizing $y$-parameter of $\sim1.4$ and a larger coronal optical depth of $\sim5$. This result, when attributed to the high inclination of EXO 0748$-$676, suggests that the Comptonizing corona may elongate along the disk plane, and give a longer path for the seed photons when viewed from edge-on inclinations.     
 
\end{abstract}

\keywords{accretion, accretion disks $-$ Stars: neutron $-$ X-rays: binaries $-$ radiative transfer}

%%%%%%%%%%%%%%%%%%%%%%%%
\section{Introduction}
\label{sec:introduction}
%%%%%%%%%%%%%%%%%%%%%%%%

Neutron-star low-mass X-ray binaries (hereafter LMXBs), luminous X-ray objects in the Milky Way and other galaxies, are low-magnetic-field neutron stars (NSs) accreting materials from their low-mass companions through Roche lobe overflow. They are found either in the high/soft state (HSS) or the low/hard state (LHS) \citep[e.g.,][]{White85,Mitsuda84,Mitsuda89}. For ``atoll" sources including the LMXBs studied in the present paper, these physical states correspond respectively to the ``banana" and ``island" states \citep{Hasinger89}, which are based on empirical color-color diagrams. In the LHS, an LMXB exhibits a rather hard spectrum extending to $\sim$ 100 keV, which is considered to arise from thermal Comptonization, wherein some soft photons are boosted to higher energies by hot electrons \citep{Sunyaev80} in a vicinity of the NS. The cloud of hot electrons is often referred to as ``Comptonizing corona" \citep[hereafter CC;][]{Chapline73,Lamb79}. 

Generally, the CC is identified as a geometrically thick and optically thin accretion flow, which is considered to form near an accreting compact object when the mass accretion rate is relatively low \citep{Lightman74}. Recent \suzaku~observations \citep[e.g.,][]{Sakurai12,Sakurai14} are confirming that the CC in several LMXBs is accreting relatively isotropically onto the NS, at least when the luminosity is $\sim1\%$ of the Eddington luminosity \citep[$L_{\rm edd}$;][]{Matsuoka13,Sakurai14}. However, the more detailed geometry of the inflowing corona is still unclear. In particular, it is still unknown whether it is very spherical, or oblate towards the disk plane. Thus we are motivated to compare ``dipping" and ``non-dipping" LMXBs in the LHS, in order to further study the coronal shape. 

Dipping LMXBs, or ``dippers", are characterized by periodic dips in their X-ray intensity, and are considered to be intrinsically the same as non-dipping LMXBs, except for their more edge-on inclination angles. If CC has an oblate shape, dipping LMXBs would show systematically stronger Comptonization in the non-dipping period, because the seed photons must pass through the CC with a longer path. Indeed, dippers have been reported to exhibit rather hard X-ray spectra over a large range of luminosity. Examples include \textit{BeppoSAX} observations of 4U 1915$-$05 \citep{Church98}, XB 1254$-$690 \citep{Iaria01}, XB 1323$-$619 \citep{Church99} and EXO 0748$-$676 \citep{Sidoli05}. 

In these previous studies of dipping LMXBs, the X-ray spectra were in most cases fitted with an empirical cutoff power-law model. However, this is not physical, and does not enable detailed comparison between dippers and other LMXBs. Given this, \citet{Zhang14} studied a \suzaku~data set of 4U 1915$-$05 in the HSS. They successfully reproduced the 0.8 $-$ 45 keV spectrum of this source, using the canonical two component spectral model \citep{Mitsuda84} that describes broadband X-ray spectra of non-dipping LMXBs in the HSS. Furthermore, they found that the blackbody component of this dipper is more strongly Comptonized than in normal LMXBs. This gives evidence that the CC in the HSS is rather flattened to the accretion disk plane, in agreement with the report by \citet{Gladstone07}, that dipping LMXBs in the HSS have systematically harder spectra than normal LMXBs. In the LHS, however, evidence of flattened coronae has so far been obtained only in black hole binaries \citep[BHBs, e.g.,][]{Makishima07,Heil15}. So, in the present paper, we intend to carry out similar studies for LMXBs in the LHS. 

For the above purpose, we again choose the Japanese satellite \suzaku~\citep{Mitsuda07}, which possesses both a high energy resolution and a broadband coverage. Among the four LHS dipping LMXBs (EXO 0748$-$676, XB 1323$-$619, XTE J1710$-$281 and 4U 1822$-$37) in {\it suzaku} archive, we selected EXO 0748$-$676, which has the highest X-ray flux as indicated in Table 1 of \citet{Zhang14}. From an {\it EXOSAT} observation, \citet{Parmar86} first discovered that EXO 0748$-$676 shows irregular intensity dips and a total eclipse, synchronized with its orbital period of 3.82 hr, and constrained the source inclination angle to be $i = 75^{\circ}$ $-$ $83^{\circ}$. We hence fix $i = 80^{\circ}$ in the present paper. Since then, the object has been studied by many authors, including \citet{Cottam01}, \citet{Jimenez03}, \citet{Sidoli05} and \citet{Trigo11}. However, except for \citet{Sidoli05}, most of these publications focused on the dip phenomenon or X-ray bursts, without much attention to the intrinsic persistent emission. In the present work, we utilize the \suzaku~observation of this source, in order to obtain physical constrains on the source accretion scheme (section \ref{sec:model}), including the shape of the CC (from section \ref{sec:shape}). 

In the present paper, we employ a source distance of $d=7.1\pm1.2$ kpc, which is based on the detailed analysis of helium-dominated X-ray bursts with photospheric radius expansion \citep{Galloway08b}. This value is consistent with an upper limit from analysis of persistent spectra by \citet{Trigo11}, but is larger than is indicated by the X-ray burst analysis assuming hydrogen-dominated atmosphere \citep[$d$ $\sim$ 5 kpc,][]{Wolff05,Galloway08a}. Thus, we include a distance uncertainty of 30\%.     

%######################
\section{Observation}
\label{sec:observation}
%######################

EXO 0748$-$676 was observed by \suzaku~on 2007 December 25 (MJD 54459) with ObsID of 402092010. The observation started at 05:41:13, and ended at 07:00:24 the next day with an elapsed time of 91.0 ks and a net exposure time of 45.9 ks. The soft X-ray band (0.2 $-$ 12 keV) was covered by three cameras of the X-ray Imaging Spectrometer (XIS 0, XIS 1 and XIS 3) on board, while XIS 2 had stopped working before this observation. We utilized only XIS 0 and XIS 3 to extract light curves (section \ref{sec:lightcurve}) and spectrum (section \ref{sec:spectra}), because these front-illuminated CCD cameras are more accurately calibrated than XIS 1 which uses a back-illuminated CCD chip. In the hard X-ray band above 10 keV, we analyzed HXD-PIN data which have an energy coverage up to 70 keV. The data from HXD-GSO were not employed because the source was not significantly detected therein. Both the XIS and HXD-PIN data were analyzed by HEAsoft version 6.13, with the calibration database of version 20070731 for the XIS, and version 20070710 for the HXD.

\subsection{XIS and HXD-PIN Data Analysis} 
\label{sec:analysis}

In the present observation, XIS 0 and XIS 3 were both operated in the normal mode using 1/4 window option. For each CCD camera, events with pixel formats of $3\times3$ and $5\times5$ were combined, and the events with grades 0,2,3,4 and 6 were selected. The total count rate was $\sim 11.0$ cts s$^{-1}$ per XIS camera. With a time resolution of 2 s, photon pileup is negligible \citep{Yamada12}. Thus, around the X-ray center at $\alpha$ = 07$^{\rm h}$48$^{\rm m}$31.19$^{\rm s}$ and $\delta$ = -67$^{\circ}$45$^{\prime}$09.52$^{\prime\prime}$ (J2000.0), we accumulated on-source events over a circle of $2^{\prime}$, which contains 90\% of the source counts. The background regions were chosen to be two circles from two edges of the CCD, each located $7^{\prime}$ from the source center, with a radius of $1^{\prime}.4$; the total background area equals to the source area. The background-subtracted source count rate in the 0.6 $-$ 10 keV range is $9.33\pm0.01$ cts s$^{-1}$, when averaged between XIS 0 and XIS 3.

The source was detected up to 55 keV with HXD-PIN. The raw 12 $-$ 55 keV count rate was $1.06\pm0.01$ cts s$^{-1}$ on average, among which $0.54$ cts s$^{-1}$ was Non X-ray Background (NXB) according to the NXB model provided by the HXD team. Contribution from the Cosmic X-ray Background (CXB) was calculated to be $0.02$ cts s$^{-1}$ using a model as 
\begin{equation}
\textmd{CXB}(E) = 9.41\times 10^{-3}  \left(\frac{E}{1~\textmd{keV}}\right)^{\hspace{-0.3em}-1.29}\hspace{-1.4em}\exp\left(-\frac{E}{40~\textmd{keV}}\right)   
\label{eq:cxb}
\end{equation}
\citep{Boldt87}, where the unit is photons cm$^{-2}$ s$^{-1}$ keV$^{-1}$ FOV$^{-1}$. After subtracting both NXB and CXB, and further correcting the data for the dead time with a command {\tt hxddtcor}, the 12 $-$ 55 keV PIN signal count rate became $0.58\pm0.01$ cts s$^{-1}$.

%#######################
\subsection{Light Curves}
\label{sec:lightcurve}
%#######################

Figure 1 (left panels) shows background-subtracted (using {\tt lcmath}) light curves of the XIS and HXD-PIN, and the ratio of the latter to the former. From the HXD-PIN data we subtracted only the NXB, since the CXB is constant and rather minor. Multiple dips are seen in the light curves, especially in that of the XIS, as repeatedly reported and explained to be caused by an ionized absorber on the accretion disk \citep{Trigo06}. Four type I X-ray bursts were also detected at observation times of 30.0 ks, 47.0 ks, 64.6 ks and 85.7 ks. Detailed analysis of the dips and bursts are out of the scope of this paper. Except for the dips and bursts, both the XIS and HXD-PIN count rates, and hence the hardness ratio, remained constant within $\sim$10\%.  

After excluding the bursts, we folded the XIS and HXD-PIN light curves and the hardness ratio at the orbital period of $P_{\rm orb}$ = 13766.8 s measured by \citet{Parmar86}. Phase 0 (=1) was set to be the middle of the eclipse, which refers to 06:12:03 of 2007 December 25 (MJD = 54459.26018). The results are shown in the right panels of figure 1. The dips (except for the eclipses) are noticed only in the XIS light curve over an orbital phase $\phi$ $\sim$ 0.6 $-$ 1.1, while the rest can be considered as non-dip phases.

%#####################################
\section{Analysis of Non-Dip Spectra}
\label{sec:spectra}
%#####################################

\subsection{Preparation of Spectra}
\label{sec:preparation}

Since the XIS and HXD-PIN light curves are both approximately constant (figure 1), we utilized the whole observation period for the non-dip spectral analysis, but excluding the orbital phases of $\phi =$ 0.6 $-$ 1.1 in reference to figure 1. The type I X-ray bursts (with the 0.6-10 keV XIS count rate $>$ 12 cts s$^{-1}$) were also removed. By further requiring the simultaneous presence of the XIS and HXD-PIN data, the remaining exposure became 16.2 ks, which is $\sim$ 35\% of the total exposure. Spectra from XIS 0 and XIS 3 were co-added; so were their responses. To analyze the HXD-PIN data, we utilized the ``XIS nominal" response file released from the HXD team, and the CXB contribution was included in a model as a fixed component of equation (\ref{eq:cxb}). We chose an energy range of 0.6 $-$ 10 keV for the XIS, and 12 $-$ 55 keV for the HXD-PIN data. To avoid the calibration uncertainties around the instrumental silicon K-edge and the gold M-edge, we excluded 1.7 $-$ 1.9 keV and 2.2 $-$ 2.4 keV energy ranges of the XIS spectrum, respectively. The instrumental Al K-edge, which is not fully accounted for by the response, was fitted by an {\tt edge} model, with the edge energy fixed at 1.56 keV. The cross normalization of HXD-PIN relative to the XIS was fixed at 1.158 \citep{Kokubun07}.  

Figure 2(a) shows the derived count spectrum, and figure 2(b) shows its $\nu F_{\nu}$ form. Thus, the spectrum extends up to $\sim$ 20 keV with a hard slope, and then turns over gradually. It can be approximated (with $\chi^2/{\nu}=295.6/142$) by the model {\tt highecut$\ast$powerlaw}, defined as
\begin{equation}\label{eq:plmodel} 
A(E)=\left\{\begin{array}{ll}
KE^{-\Gamma}{\rm exp}[(E_{\rm c}-E)/E_{\rm f}] & \mbox{($E \ge E_{\rm c}$)} \\
\\
KE^{-\Gamma} & \mbox{($E \le E_{\rm c}$)} 
\end{array}\right.,
\end{equation}
with a photon index $\Gamma$ $\sim$ 1.7, a cutoff energy $E_{\rm c}$ $\sim$ 22 keV, and an e-folding energy $E_{\rm f}$ $\sim$ 60 keV, while $K$ is the normalization. The low $\Gamma$ value indicates that the source was in the LHS with strong Comptonization, while the value of $E_{\rm c}$ suggests a relatively low coronal electron temperature. The data-to-model ratio in panel (c) reveals a soft excess below $\sim$ 1 keV, which suggests that the overall emission may contain more than a single component.

\subsection{Single-Seed Comptonization Models} 
\label{sec:single-seed}

Although the empirical model in figure 2 is roughly successful, it is not yet fully acceptable, and it has little physical meaning. Hence, we attempt to fit the spectrum with the canonical two-component spectral model for LMXBs, {\tt diskbb+bbody} \citep{Mitsuda84}, applying the Comptonization code {\tt nthcomp} \citep{Zdziarski96,Zycki99} on either (single-seed) or both (double-seed) of the two optically-thick components. In \citet{Sakurai12}, the model {\tt diskbb+nthcomp[bbody]}, in which the seed photons are from the NS surface as indicated by the square bracket, successfully explained the LHS spectra of Aql X$-$1 obtained with \suzaku. Thus, we first applied the same model to the spectra of EXO 0748$-$676. Column density of photoelectric absorption was fixed to the Galactic line-of-sight value of $N_{\rm H} = 1.1\times10^{21}$ cm$^{-2}$, because the source locates $\sim2.4$ kpc below the Galactic plane, as calculated from its Galactic latitude $b=-19.8^{^\circ}$, and the source distance of 7.1 kpc; hence, the source is outside the neutral hydrogen disk of the Galaxy. 

The obtained results are shown in figure 3(a). Around 0.8 $-$ 1.2 keV, we noticed some negative residual features that cannot be explained by the Galactic absorption, with the fit goodness of $\chi^2(\nu)$ = 175.0 (142). Thus we added a partial ionized absorber with the code {\tt zxipcf} in {\tt XSPEC}, using a grid of XSTAR photoionized absorption models \citep{Reeves08}. The absorber column density $N_{\rm abs}$, the ionization parameter $\log(\xi)$, and the absorber covering fraction were set to be free, while the redshift was fixed to 0. The fit gave $N_{\rm abs}~<~1.1\times10^{22}$ cm$^{-2}$ and $\log(\xi)~\sim~2.4$, together with $\chi^2(\nu)$ = 166.5 (139). However, as shown in figure 3(b), an absorption feature was still evident at $\sim$ 1 keV, which may be related to Neon Ly$\alpha_{1}$. We hence added a Gaussian absorption model, with the $1\sigma$ line width fixed at 20 eV. 

Through the above improvements, the fit has become fully acceptable with $\chi^2(\nu)=142.8(137)$, as shown in table \ref{tab:fitting} labeled as ``single-seed" model, and in figure 3(c). The obtained inner-disk temperature \ktin~$\sim$ 0.2 keV and the radius \rin~$\sim$ 160 km indicate a disk truncated at a large radius. For the NS blackbody, which serves as the seed component of the {\tt nthcomp} model, the data gave only an upper limit of \ktbb~$\sim$ 0.25 keV, and a lower limit of \rbb~$\sim$ 33 km as specified by the {\tt nthcomp} normalization. However, this value of \rbb~is much larger than the neutron star radius, even considering the 30\% distance uncertainty. This means that the data requires more seed photons than are available from the NS surface, encouraging us to consider a ``double-seed" modeling in the next subsection. 

In addition to the above attempts, we also considered two more single-seed Comptonization models. One is {\tt bbody+nthcomp[diskbb]}, assuming that the seed photons are supplied only from the disk. This is equivalent to some disk Comptonization models \citep[e.g.,][]{Church95}. The other is a partial NS Comptonization model, {\tt bbody+nthcomp[bbody]}, in which some fraction of the NS blackbody is Comptonized while the rest is directly visible, and the disk emission is not detected. Both models gave acceptable fits to the data; however, neither of them was physically acceptable. The disk Comptonization model gave too large a blackbody radius as \rbb~$>$ 100 km, trying to explain the soft excess with a rather low temperature, and too large a disk radius as \rin~$>$ 1000 km to supply a sufficient number of seed photons. In the partial NS Comptonization model, the directly seen {\tt bbody} component and the Comptonized seed {\tt bbody} both became too large in radius as \rbb~$>$ 100 km, for a similar reason as above. Thus, we no longer consider these two models hereafter. 

We further examined the spectrum for the possible presence of other additional components, especially a directly seen NS blackbody emission \citep[e.g.,][]{Lyu14}, and disk reflection \citep[e.g.,][]{DiSalvo15}. By adding another {\tt bbody} component to our final single-seed solution (figure 3c), the fit was improved by $\Delta \chi^2= -7.46$ for $\Delta \nu$ = -2. The derived blackbody temperature is $0.4^{+0.1}_{-0.2}$ keV with a reasonable blackbody radius of $2.9^{+8.0}_{-0.9}$ km. However, the Comptonized {\tt bbody} component did not change significantly, so that the total blackbody radius further increased and hence the fit became even more unphysical. For the disk reflection, we multiplied a {\tt reflect} component on our model and derived a scaling factor (the solid angle divided by 2$\pi$) of $\sim$ 0.16. However, the fit was improved only by $\Delta \chi^2= -0.9$ for $\Delta \nu$ = -1; thus we do not consider that a reflection component is very necessary in this high inclination system.

\subsection{Double-Seed Comptonization Models} 
\label{sec:double-seed}

In order to overcome the seed-photon shortage revealed in section \ref{sec:single-seed}, we next consider a ``double-seed" Comptonization, namely, the case wherein the disk emission and the NS blackbody are both Comptonized to jointly produce the hard continuum. The model thus becomes {\tt nthcomp[diskbb]+nthcomp[bbody]} with the common Galactic absorption. The values of \kte~and $\Gamma$ were first constrained to be the same between the two Comptonization components, but the fit was far from acceptable since the model could not explain the slightly concave spectral shape. As indicated by recent results on LMXBs in the LHS \citep[e.g.,][]{Sakurai14}, the disk is likely to be truncated at a large distance from the central NS, and hence it is reasonable to assume different optical depths for the two Comptonization components. Hence, we allowed them to have different values of $\Gamma$, while \kte~was kept the same. This model was found to be successful with $\chi^2(\nu)$ = 134.6 (136). Compared to the single-seed model, it is better by $\Delta\chi^2=-6.5$, for a decrease of only by $\Delta\nu=-1$. 

The successful fit obtained above is shown in table \ref{tab:fitting} as ``double-seed" model, and in figure 3(d). Thus, the Comptonized disk emission takes a larger fraction of the soft-band emission below 2 keV, compared to the solution of the single-seed model. As a result, the NS component moved to higher energies with a higher \ktbb~and a considerably smaller \rbb~which is physically acceptable within errors. This solves the problem encountered in the single-seed modeling. The value of \rin~has become even larger than that obtained in the single-seed modeling, but its lower limit is still reasonable at $\sim$ 150 km, calculated using the lower limit on the distance \citep[5 kpc,][]{Galloway08a} and the upper limit on the inclination \citep[$75^{\circ}$,][]{Parmar86}. The coronal temperature has remained relatively unchanged from the single-seed model. The slope $\Gamma$ for the disk has been found to be much larger (steeper) than that for the NS, resulting in a $\sim$ 5 times smaller coronal optical depth (table 1). This is reasonable considering the large value of \rin. We further allowed the two Comptonized components to have different values of \kte. Then, the disk component favored a lower value of \kte, but the fit goodness did not change significantly. 

Based on this modeling, the unabsorbed source flux in 0.6 $-$ 55 keV was calculated to be $5.7\times10^{-10}$ erg cm$^{-2}$ s$^{-1}$, which gives an unabsorbed luminosity of $3.4\times10^{36}$ erg cm$^{-2}$ s$^{-1}$ in 0.6 $-$ 55 keV and $5.2\times10^{36}$ erg cm$^{-2}$ s$^{-1}$ after bolometric correction. The latter corresponds to $\sim$ 0.025 $L_{\rm edd}$ assuming a hydrogen-dominated donor star. The luminosity could be twice higher if we considered the inclination effect, or lower by a factor of 2 if taking into account the 30\% distance uncertainty.

%#####################################################
\section{\suzaku~Archived Sample of Non-dipping LMXBs}
\label{sec:sample}
%#####################################################   

So far, we have successfully explained the \suzaku~spectra of EXO 0748$-$676 using a spectral model that was developed to explain those of non-dipping LMXBs in the LHS. The next task is to examine whether the derived model parameters, in particular $\tau$ and \kte, differ between the present dipping source and other non-dipping ones. For this purpose, we performed, after \citet{Sakurai15}, a survey of ``normal" LMXB spectra in the \suzaku~archive. In addition to dippers, we also excluded high magnetic field sources \citep[e.g., Her X$-$1, GX 1+4, 4U 1626$-$67, and 4U 1822$-$37;][]{Sasano14}, ultra compact X-ray binaries (with hydrogen-depleted donor stars), and symbiotic binaries (with evolved companions), because the accreting matters in the latter two types of systems are considered to have non-solar electron-to-baryon number ratios, which may affect the Comptonization details. As listed in Table \ref{tab:allsample}, this selection left us with 43 observations of 16 normal LMXBs. As shown by \citet{Sakurai15}, the hardness ratio $H$ between the 20 $-$ 40 keV HXD-PIN signal rates to the 5 $-$ 10 keV XIS-FI signal rates (both with the background subtracted) serves as a good state indicator. Thus, 16 observations of 8 sources were found to sample the LHS, with $H > 0.03$; these are listed in Table \ref{tab:sampleresult}. Among them, the six observations of Aql X$-$1 (ObsID from 402053020 to 402053070) were previously analyzed by \citet{Sakurai14} and were updated by \citet{Sakurai15}. Besides, GS 1826$-$238 was further studied in detail by \citet{Ono16}, of which the results are quoted in the present paper.

We analyzed the data of the remaining 15 observations employed similar procedures as described in section \ref{sec:analysis}. A major difference is that the XIS background events were taken from an annulus with the inner radius of $4^{\prime}$ and the outer radius of $5^{\prime}$. Moreover, the emission from Aql X$-$1 was detected with HXD-GSO up to $> 100$ keV. 

We then fit the spectra of the XIS and the HXD jointly, using either the single-seed or the double-seed models employed in the previous section. As for the Comptonization code, either {\tt nthcomp} or {\tt compPS} \citep{Poutanen96} was adopted when the optical depth is $\tau \gtrsim 2$ or $\tau \lesssim 2$, respectively. The latter model was employed under spherical geometry, namely the geometry parameter of 4 in XSPEC. In an overlapping parameter region at $\tau \sim 2$, the two codes yielded very similar \kte~and $\tau$ within $\sim 3\%$ \citep{Sakurai15}. Since these LHS sources span a large range ($\sim$ 5 orders of magnitude) of luminosities (or mass accretion rates), their spectra exhibited some variety. For example, the disk emission was undetectable from the two faintest sources, Cen X$-$4 and SLX 1737$-$282, and the three faintest observations of Aql X$-$1. Twelve out of the 15 spectra were explained by the single-seed modeling (section \ref{sec:single-seed}), while the remaining three (two from 4U 1705$-$44, and one from Aql X$-$1) required the double-seed modeling. However, unlike EXO 0748$-$676, these data sets preferred a common corona with the same $kT_{\rm e}$ and $\tau$ for the two-seed sources. The disk reflection was found to be necessary only for three observations of Aql X-1 (ObsID from 402053020 to 402053040), as already published in \citet{Sakurai12}. The fit results to the 15 spectra are summarized in Table \ref{tab:sampleresult}. We discuss these results in section \ref{sec:discussion}.

%##################################
\section{Discussion and Conclusion}
\label{sec:discussion}
%##################################

\subsection{The Spectral Modeling of EXO 0748$-$676}
\label{sec:model}

During the present \suzaku~observation, EXO 0748$-$676 was in the LHS because it exhibited a very hard spectral slope of $\Gamma \sim 1.7$ up to $\sim$ 20 keV (figure 2). This classification is also supported by its high value of the hardness ratio (section \ref{sec:sample}), $H$ = $0.077\pm0.002$. Its persistent luminosity, $L_{\rm bol} \sim 0.025^{+0.025}_{-0.013}~L_{\rm edd}$, is relatively high among LMXBs in the LHS, considering that the soft-to-hard state transition luminosity for LMXBs is generally 0.01 $-$ 0.04 $L_{\rm edd}$ \citep{Maccarone03,Matsuoka13}. Such a luminous hard-state LMXB is however not rare, as the luminosity range of the \suzaku~LHS sample spans from $\lesssim$ 0.01 to $\sim$ 0.1 $L_{\rm edd}$ (Table \ref{tab:sampleresult}). 
 
The broadband \suzaku~spectrum of this object has been described successfully with the canonical two-component model for LMXBs \citep{Mitsuda84}, namely, the NS and disk emissions, plus Comptonization. The data indicate a ``double-seed" Comptonization condition (sections \ref{sec:single-seed} and \ref{sec:double-seed}), where the NS blackbody emission and the disk emission are both Comptonized, to form the hard continuum and to acquire a higher color temperature, respectively. Assuming a common electron temperature, the obtained value of \kte~$\sim 13$ keV is relatively low for LMXBs in the LHS, as judged from the compilation in Table \ref{tab:sampleresult}, where most of the other sources show \kte~$>$ 20 keV. The optical depths of the two components were calculated using equation (2) of \citet{Zhang14}, so were the Comptonizing $y$-parameters defined therein. As clearly seen in figure 3(d), the disk component is less Comptonized than the NS component, because the corresponding disk/NS ratio of $\tau$ is $\sim$ 1/5, and the ratio of $y$-parameters is $\sim$ 1/10. Considering the large value of \rin, it suggests that the corona is centrally localized, and the disk intrudes into only the outer part of the corona. 
Similar results were obtained by \citet{Ono16} on GS 1826$-$238, that only an inner region of a truncated accretion disk is Comptonized.

We further compared the above results with the compilation of the \suzaku~sample shown in Table \ref{tab:sampleresult}. There, the fit results to the 16 LHS spectra of the 8 LMXBs can be classified into three categories, based on their best-fit models; while these categories all involve Comptonized NS emission, their difference is in the disk emission. The five observations in category 1 (C1) did not have disk emission detected in the \suzaku~spectra, and their luminosities are all $< 0.01 L_{\rm edd}$. In seven observations with $L_{\rm X} \gtrsim 0.01 L_{\rm edd}$, forming category 2 (C2), the disk was detected but it did not need to be Comptonized. Finally, in the four most luminous observations with $L_{\rm X} \sim 0.1 L_{\rm edd}$, to be called category 3 (C3), the disk emission was also strongly Comptonized by a corona that can be considered either to be the same (with the same \kte~and $\tau$ values) as for the NS Comptonization (406010010 of Aql X$-$1 and 4U 1705$-$44), or to have a lower temperature \citep[GS 1826$-$238;][]{Ono16}. The luminosity of EXO 0748$-$676 is higher than in most of the LMXBs in C2, and lower than (ignoring the inclination effect) in all LMXBs in C3. The accretion scheme of this source is likely in between those of LMXBs with single-seed and double-seed Comptonization, because the fit prefers disk Comptonization, but to a less extent than that of the NS. 

Further comparing the results on EXO 0748$-$676 (Table \ref{tab:fitting}) with those of the other LMXBs (Table \ref{tab:sampleresult}), two possible differences are noticed; the considerably larger value of \rin, and the somewhat lower \ktbb~($\lesssim$ 0.4 keV vs. $\gtrsim$ 0.5 keV). However, we consider these to be still within systematic uncertainties of the spectral modeling, for the following reasons. First, the disk emission of EXO 0748$-$676 is very much weakened by the high inclination, and couples strongly with the neutral and ionized absorptions at the softest end of the spectrum. As a result, the {\tt diskbb} and its Comptonization must be subject to considerable systematic uncertainties. Second, in figure 3(d), the value of \ktbb~is constrained only by the 1$-$2 keV data points, where the Rayleigh-Jeans regime of the underlying blackbody is directly visible. Since this is the very energy range where the disk emission and the Comptonized blackbody overlap, the systematic uncertainties will propagate from the former to the latter. To test the above inferences, we re-fitted the EXO 0748$-$676 spectra with the same model as in figure 3(d), but fixing \ktbb~at 0.5 keV. Then, the fit goodness decreased only slightly, by $\Delta \chi^2= 16.2$ for $\Delta \nu$ = $+1$. Therefore, we do not consider that the model parameters of EXO 0748$-$676 are deviated from those of the other objects.

\subsection{Properties of the Comptonizing Corona}
\label{sec:shape}

Now we discuss the Comptonizing effects of EXO 0748$-$676. Judging directly from the spectral shape, EXO 0748$-$676 exhibited a hard spectral slope of $\Gamma=1.7$ in 5 $-$ 20 keV, an energy band dominated by the Comptonized NS emission and is free from interstellar absorption. It could be even harder as once revealed by $BeppoSAX$ in 2000, wherein the spectrum extended to $\sim$ 100 keV with $\Gamma \sim 1.3$ \citep{Sidoli05}. Other non-dipping LHS LMXBs, in contrast, show a rather flat slope in $\nu F_{\nu}$ plot of $\Gamma \sim 2$ in the same energy range. Examples in the literatures include the persistent LMXBs in the LHS, GS 1826$-$238 \citep[figure 2 of][]{Ono16}, 4U 1728$-$34 \citep[figure 6 of][]{Tarana11}, 4U 1705$-$44 \citep[top panel in figure 3 of][]{Lin10} and 4U 1812$-$12 \citep[figure 2 of][]{Tarana06}, and the transient LMXBs in the LHS, Aql X$-$1 \citep[figure 2 of][]{Sakurai14} and 4U 1608$-$522 \citep[top left panel in figure 5 of][]{Tarana08}. The harder spectral slope indicates that EXO 0748$-$676 has stronger Comptonization of the NS component than the others.   

Let us more quantitatively compare EXO 0748$-$676 with the \suzaku~sample. For this purpose, figure 4(a) shows the calculated Compton $y$-parameter of the present sources, including EXO 0748$-$676, against their bolometric luminosity $L_{\rm X}$. From the low luminosity end, the $y$-parameter remains approximately constant at $\sim$ 0.5 within errors till $2 - 4 \times 10^{36}$ erg s$^{-1}$, and then increases towards higher luminosities. The point of EXO 0748$-$676, plotted in diamond, is an obvious outlier with a $\sim$ 1.5 $-$ 3 times higher $y$ value than sources with similar $L_{\rm X}$. This confirms that the NS emission is more strongly Comptonized in EXO 0748$-$676 than in the other sources. Furthermore, this effect must be attributed to a larger $\tau$ of EXO 0748$-$676, because it has a rather low \kte. Indeed, this expectation is directly confirmed in an $L_{\rm X}$-$\tau$ plot presented in figure 4(b), where EXO 0748$-$676 again stands out.       

Although the luminosity is the primary quantity to control the behavior of LMXBs, it is subject to distance (and some extent inclination) uncertainties. Moreover, a hysteresis effect \citep{Meyer05,Miyamoto95} may complicate the behavior of sources near the transition luminosities, including EXO 0748$-$676. Hence, we introduce a new parameter $Q \equiv kT_{\rm e}/kT_{\rm bb}$ proposed by \citet{Makishima14}, which represents a balance between the electron heating from ions, and their cooling through Comptonization by the seed photons. This quantity is independent of the source distance, and has been confirmed to be free from the hysteresis. It hence serves as a good state indicator, which separates the HSS ($Q\lesssim7$) from the LHS ($Q\gtrsim10$) \citep{Makishima14,Sakurai15}. We show $Q$-$y$ and $Q$-$\tau$ diagrams in figure 4(c) and figure 4(d), respectively. Thus, the systematically larger $y$ and $\tau$ of EXO 0748$-$676 are still apparent, even when $L_{\rm X}$ is replaced by $Q$. 

Because EXO 0748$-$676 has a very high inclination compared to the other LMXBs considered here, its large $\tau$ can be most naturally ascribed to a flattened coronal shape, which will give a longer path length when viewed from high inclinations. As indicated by figure 4(b) and 4(d), the $\tau$ value of EXO 0748$-$676 is about twice of the average value of the other LMXBs with the same $L_{\rm X}$ and $Q$. Assuming that the corona has an axi-symmetric ellipsoidal geometry, that the electron density does not depend significantly on the sources, and that the inclination is $i=80^{\circ}$ in EXO 0748$-$676 and typically $\sim$ $45^{\circ}$ in the other sources, the higher value of $\tau$ by a factor of $\sim$ 2 can be explained if the corona has an aspect ratio of $\sim$ 3:1. Combined with \citet{Zhang14}, who found evidence for a flattened corona in the HSS dipper 4U 1915-05, the present study further suggests that the CC in the LMXBs has an oblate shape in general, although its exact geometry is yet to be quantified. 

The present results are very similar to the reported evidence for flattened coronae in BHBs in the LHS by \citet{Makishima07} and \citet{Heil15}. Especially, \citet{Heil15} reported that flattened coronae are present in a wide range of hard and intermediate states. In summary, we suggest that the flattened coronal shape is common to LMXBs and BHBs, and to be rather independent of their spectral states.

%###############
\acknowledgments
%###############

This work was supported by the Japan Society for the Promotion of Science (JSPS) under the Grant-In-Aid number 24-02321, the Ministry of Science and Technology of China (grant No. 2013CB837900), and the National Science Foundation of China (grant Nos. 11125313 and 11433002).

\newpage
%##########
%REFERENCES
%##########

%%%%%%%%%%%%%%%%%%%%%%%%%%%%%%%%%%%%%%%%%%%%%%%%
\clearpage

%=======
%Table 1
%=======

\begin{table}
\begin{center}
\caption{Results of the model fittings to the 0.6 $-$ 55 keV \suzaku~spectra of the persistent emission of EXO 0748$-$676.$^a$}
\label{tab:fitting}
\vspace{0.3cm}
\begin{tabular}{cccc}
\hline
\hline
Component   &  Parameter                                    &	   Value		    &	      \\
\hline
model       &                                               &  single-seed\tablenotemark{b} &  double-seed\tablenotemark{c}	\\	 
\hline
wabs        &  $N_{\rm H}(10^{22}$ cm$^{-2})$               &  0.11 (fixed)		    &  0.11 (fixed)		\\ 
\hline
zxipcf  & $N_{\rm abs}(10^{22}$ cm$^{-2})$ & $0.14^{+0.95}_{-0.09}$  &  $0.38(<5.85)$    \\ 
        &  $\log(\xi)$      &	$2.36^{+0.29}_{-0.26}$   &  $2.34^{+0.78}_{-0.73}$   \\ 
        &  CvrFract         &	$0.59(>0.14)$		 &  $0.21(<0.34)$   \\  
\hline   
Al K-edge\tablenotemark{d}  &  optical depth             &  $0.03\pm0.02$  &  $0.05\pm0.02$	     \\

\hline
gabs\tablenotemark{e}       &  $E$ (keV)  &	$1.03\pm0.02$		             &  $1.03\pm0.02$	     \\
	                    &  line strength ($10^{-3}$)    &	$5.6\pm2.0$	          &  $5.0\pm2.0$   \\

\hline 
{\tt diskbb}  &  $kT_{\rm in}$ (keV)	                    &	$0.17\pm0.01$		                  &  $0.14\pm0.01$	      \\  
              &  $R_{\rm in}^{f}$ (km)                      &	$159^{+26}_{-20}$        &  $284^{+28}_{-14}$ \\   
\hline 
{\tt bbody}   &  $kT_{\rm bb}$ (keV)	                    &	$0.19(< 0.25)$  	 &  $0.32^{+0.07}_{-0.09}$  \\ 
              &  $R_{\rm bb}$ (km)	                    &	$61 (> 33)$		  &  $23^{+31}_{-9}$  \\ 
\hline
{\tt nthcomp} &  seed component 	                    &   {\tt bbody}		                  &  {\tt diskbb/bbody}	\\ 
              & $\Gamma$                                    &	$1.76\pm0.01$		  &  $5.11^{+4.50}_{-0.86}$/$1.75\pm0.01$ \\  
              & $kT_{\rm e}$ (keV)                          &	$13.9^{+3.7}_{-2.0}$	  &  $12.2^{+2.4}_{-1.5}$ (common) \\ 
              & $\tau$                                      &	$4.9^{+0.5}_{-0.7}$	  &  $1.1^{+0.5}_{-0.8}$/$5.3^{+1.1}_{-0.9}$ \\ 
              & $y$-parameter                               &	$1.38\pm0.02$            &  $0.13^{+0.06}_{-0.09}$/$1.39\pm0.04$ \\
\hline      
Fit goodness  &  $\chi^2_{\nu}$(dof)                        &	1.03 (137)    &  0.99 (136)   \\ 
\hline  
\end{tabular}
\end{center}
\tablenotemark{a} The 0.6 $-$ 10 keV XIS spectrum and 12 $-$ 55 keV HXD-PIN spectrum are fitted simultaneously. The quoted errors refer to 90\% confidence, statistical only.\\
\tablenotemark{b} {\tt diskbb+nthcomp[bbody]}.\\
\tablenotemark{c} {\tt nthcomp[diskbb]+nthcomp[bbody]}, with the two Comptonized components having a common $kT_{\rm e}$ but different $\tau$.\\
\tablenotemark{d} The edge energy is fixed to 1.56 keV. \\
\tablenotemark{e} The 1$\sigma$ line width is fixed to 20 eV. \\
\tablenotemark{f} $R_{\rm in}$ is corrected for the inclination factor $\sqrt{\cos(i)}$, assuming $i=80^{\circ}$. \\
\end{table}

\clearpage

%========
%Table 2
%========

\begin{longtable}{lcccccc}
\caption{\suzaku~observations of normal non-dipping LMXBs.}\\
\hline
\hline
Name        &  OBSID    & Exp.\tablenotemark{a} & $L_{\rm abs}$\tablenotemark{b}  & $H$\tablenotemark{c} &   $D$     & Ref.\tablenotemark{d} \\
            &           & (ks)                  & (erg s$^{-1}$)                  &                      &  (kpc)    &      \\ 
\hline
\endfirsthead
\caption{continued.}\\
\hline
\hline
Name        &  OBSID    & Exp.\tablenotemark{a} & $L_{\rm abs}$\tablenotemark{b}  & $H$\tablenotemark{c} &   $D$     & Ref.\tablenotemark{d} \\
            &           & (ks)                  & (erg s$^{-1}$)                  &                      &  (kpc)    &      \\   
\hline
\endhead
\hline
\endfoot
\label{tab:allsample}

4U 1608$-$52  & 404044010 & 28.0 & 1.4$\times10^{37}$  & $(4.65 \pm 0.04 )\times 10^{-3}$   & $3.6$ & 1  \\
            & 404044020 & 25.2 & 7.7$\times10^{36}$  & $(5.38 \pm 0.07 )\times 10^{-3}$   & &    \\
            & 404044030 & 14.2 & 3.3$\times10^{36}$  & $(3.97 \pm 0.04 )\times 10^{-2}$   & &    \\
            & 404044040 & 14.5 & 1.0$\times10^{36}$  & $(5.19 \pm 0.09 )\times 10^{-2}$   & &    \\
\hline
4U 1636$-$536 & 401050010 & 20.3 & 6.5$\times10^{36}$  & $(5.18 \pm 0.05 )\times 10^{-2}$   & $5.92$ & 2 \\
            & 401050020 & 33.2 & 1.1$\times10^{37}$  & $(5.85 \pm 0.08 )\times 10^{-3}$   & &  \\
            & 401050030 & 45.2 & 8.7$\times10^{36}$  & $(1.64 \pm 0.01 )\times 10^{-2}$   & &  \\
            & 401050040 & 26.1 & 8.5$\times10^{36}$  & $(6.6 \pm 0.1 )\times 10^{-3}$     & &  \\
            & 401050050 & 11.0 & 8.6$\times10^{36}$  & $(5.0 \pm 0.1 )\times 10^{-3}$     & &  \\
\hline
4U 1705$-$44  & 401046010 & 14.4 & 8.1$\times10^{36}$  & $(4.59 \pm 0.05 )\times 10^{-2}$   & $7.4_{-1.1}^{+0.8}$ & 3 \\
            & 401046020 & 14.7 & 1.7$\times10^{37}$  & $(2.46 \pm 0.06 )\times 10^{-3}$   & &  \\
            & 401046030 & 16.7 & 6.7$\times10^{36}$  & $(2.14 \pm 0.08 )\times 10^{-3}$   & &  \\
            & 402051010 & 8.8  & 5.7$\times10^{37}$  & $(3.48 \pm 0.06 )\times 10^{-3}$   & &  \\
            & 402051020 & 14.8 & 3.7$\times10^{37}$  & $(3.44 \pm 0.06 )\times 10^{-3}$   & &  \\
            & 402051030 & 19.5 & 1.5$\times10^{37}$  & $(4.16 \pm 0.09 )\times 10^{-3}$   & &  \\
            & 402051040 & 12.4 & 4.0$\times10^{37}$  & $(3.60 \pm 0.06 )\times 10^{-3}$   & &  \\
            & 406076010 & 84.5 & 4.6$\times10^{36}$  & $(6.10 \pm 0.04 )\times 10^{-2}$   & &  \\
\hline
4U 1728$-$34  & 405048010 & 88.3 & 5.9$\times10^{36}$  & $(1.77 \pm 0.01 )\times 10^{-2}$   & $5.2$ & 4 \\
\hline
4U 1812$-$12  & 406008010 & 49.9 & 1.4$\times10^{36}$  & $(1.208 \pm 0.009 )\times 10^{-1}$ & $4$   & 5 \\
\hline
Aql X$-$1   & 402053010 & 12.5 & 9.6$\times10^{36}$  & $(2.06 \pm 0.08 )\times 10^{-3}$   & $5.2\pm0.7$ & 6 \\
            & 402053020 & 11.2 & 1.9$\times10^{36}$  & $(7.2 \pm 0.1 )\times 10^{-2}$     & & \\
            & 402053030 & 16.1 & 2.3$\times10^{36}$  & $(6.6 \pm 0.1 )\times 10^{-2}$     & & \\ 
            & 402053040 & 16.3 & 1.8$\times10^{36}$  & $(7.1 \pm 0.1 )\times 10^{-2}$     & & \\ 
            & 402053050 & 15.9 & 3.5$\times10^{35}$  & $(7.1 \pm 0.3 )\times 10^{-2}$     & & \\ 
            & 402053060 & 19.8 & 9.1$\times10^{33}$  & $(6 \pm 2 )\times 10^{-2}$         & & \\
            & 402053070 & 13.4 & 1.1$\times10^{34}$  & $(1.2 \pm 0.3 )\times 10^{-1}$     & & \\
            & 406010010 & 32.8 & 1.0$\times10^{37}$  & $(6.89 \pm 0.03 )\times 10^{-2}$   & & \\
            & 406010020 & 35.2 & (2.7-3.7)$\times10^{37}$  & $0.12 - 0.008$               & & \\
            & 406010030 & 34.0 & 4.7$\times10^{37}$  & $(3.08 \pm 0.02 )\times 10^{-3}$   & & \\
\hline
Cen X$-$4   & 403057010 & 124.0& 5.1$\times10^{31}$  & $(9 \pm 3 )\times 10^{-2}$         & $1.2\pm0.3$ & 7\\ 
\hline
Cyg X$-$2     & 401049010 & 34.5 & 5.7$\times10^{37}$  & $(2.15 \pm 0.02 )\times 10^{-3}$   & $7.28$ & 8\\
            & 403063010 & 81.9 & 5.9$\times10^{37}$  & $(6.50 \pm 0.04 )\times 10^{-3}$   & & \\
\hline
GS 1826$-$238 & 404007010 & 82.0 & 1.1$\times10^{37}$  & $(1.038 \pm 0.004 )\times 10^{-1}$ & $7_{-3}^{+1}$ & 9 \\
\hline
GX 17+2     & 402050010 & 15.0 & 1.5$\times10^{38}$  & $(6.23 \pm 0.05 )\times 10^{-3}$   & $9.8$ & 10\\
            & 402050020 & 17.7 & 7.2$\times10^{37}$  & $(1.92 \pm 0.01 )\times 10^{-2}$   & & \\
            & 406070010 & 81.1 & 1.8$\times10^{38}$  & $(6.28 \pm 0.02 )\times 10^{-3}$   & & \\
\hline
GX 340+0    & 403060010 & 80.3 & 8.5$\times10^{37}$  & $(9.05 \pm 0.04 )\times 10^{-3}$   & $11\pm3$ & 11 \\
\hline
GX 349+2    & 400003010 & 18.7 & 4.5$\times10^{37}$  & $(3.82 \pm 0.03 )\times 10^{-3}$   & $5$ & 12 \\ 
            & 400003020 & 23.9 & 1.9$\times10^{37}$  & $(2.94 \pm 0.02 )\times 10^{-2}$   & & \\
\hline
GX 9+9      & 404071010 & 57.6 & 7.6$\times10^{37}$  & $(4.57 \pm 0.04 )\times 10^{-3}$   & $10$ & 13 \\
\hline
LMC X$-$2     & 401012010 & 69.0 & 1.5$\times10^{38}$  & $(1.01 \pm 0.05 )\times 10^{-3}$ & $50$ & 14 \\
\hline
Ser X$-$1     & 401048010 & 15.3 & 4.8$\times10^{37}$  & $(2.82 \pm 0.03 )\times 10^{-3}$ & $8.4$ & 12 \\
\hline
SLX 1737$-$282& 503103010 & 29.0 & 3.5$\times10^{35}$  & $(8.2 \pm 0.4 )\times 10^{-2}$   & $5-8$ & 15 \\
\hline
\end{longtable}
\tablenotemark{a} Net exposure per XIS sensor. \\
\tablenotemark{b} Absorbed luminosity in $0.8-60$ keV. For SLX 1737$-$282, the 6.5 kpc median value of the source distance was used. \\
\tablenotemark{c} Hardness of the 20 $-$ 40 keV HXD-PIN signal rates to the 5 $-$ 10 keV XIS-FI signal rates. \\
\tablenotemark{d} References from which the source distance is quoted:
$^1$\citet{Nakamura89}, $^2$\citet{Cornelisse03}, $^3$\citet{Haberl95}, $^4$\citet{Galloway03}, $^5$\citet{Cocchi00}, $^6$\citet{Jonker04}, $^7$\citet{Chevalier89}, $^8$\citet{Orosz99}, $^9$\citet{Barret00}, $^{10}$\citet{Kuulkers02}, $^{11}$\citet{Fender00}, $^{12}$\citet{Christian97}, $^{13}$\citet{Savolainen09}, $^{14}$\citet{Feast99}, $^{15}$\citet{intZand02}.\\

%========
%Table 3
%========
\clearpage
\setlength{\headsep}{30mm}
\begin{landscape}
\begin{table}
\caption{The model fittings to the broadband \suzaku~continuum spectra of the LHS LMXBs.}
\scriptsize
\label{tab:sampleresult}
\vspace{0.3cm}
\begin{tabular}{lcccccccccccc}
\hline
\hline
Name   & ObsID  & Model & $L_{\rm X}$\tablenotemark{a}  & $N_{\rm H}$	   & $kT_{\rm in}$ & $R_{\rm in}$\tablenotemark{b}  & $kT_{\rm bb}$  & $R_{\rm bb}$ & $kT_{\rm e}$ & $\tau$ & $y$    & $\chi^2_{\nu}$(dof) \\  
       &        & 	&  (erg s$^{-1}$/$L_{\rm edd}$) & ($10^{22}$ cm$^{-2}$) &  (keV)   &  (km)			    &  (keV)	     &  (km)	    & (keV)	   &	    &    &		 \\    
\hline
\multicolumn{13}{c}{(C1) Single-seed Comptonization model without detectable disk} \\
\hline
Aql X$-$1     & 402053050  & {\tt compPS[bbody]}         & $6.48\times10^{35}/3\times10^{-3}$  & 0.36(fixed)    &   --			   & -- 	       & $0.40\pm0.01$ & $7\pm1$  & $146\pm10$ & $0.49\pm0.02$                             & $0.67\pm0.06$ & 1.05(179) \\
	      & 402053060  & {\tt compPS[bbody]}         & $1.63\times10^{34}/8\times10^{-5}$  & 0.36(fixed)    &   --			   & -- 	       & $0.27\pm0.01$ & $3\pm1$  & $170^{+50}_{-40}$	& $0.19^{+0.04}_{-0.03}$           & $0.27^{+0.08}_{-0.10}$ & 0.94(134) \\
	      & 402053070  & {\tt compPS[bbody]}         & $3.34\times10^{34}/1.6\times10^{-4}$  & 0.36(fixed)    &   --			   & -- 	       & $0.27\pm0.01$ & $3\pm1$  & $400^{+250}_{-130}$ & $0.16^{+0.03}_{-0.02}$           & $0.53^{+0.34}_{-0.22}$ & 0.81(134) \\
SLX 1737$-$282& 503103010  & {\tt compPS[bbody]}         & $6.14\times10^{35}/3\times10^{-3}$  & $1.4\pm0.1$    &   --			   & -- 	       & $0.50\pm0.04$ & $4\pm1$ & $140\pm30$ & $0.52^{+0.07}_{-0.05}$                     & $0.67^{+0.27}_{-0.20}$ & 1.02(352) \\
Cen~X-4     & 403057010  & {\tt compPS[bbody]}           & $1.08\times10^{32}/5\times10^{-7}$  & 0.05(fixed)    &   --	           & -- 	       & $0.19\pm0.01$ & $0.5\pm0.2$ & $400^{+700}_{-200}$ & $0.11^{+0.07}_{-0.02}$        & $0.36^{+0.42}_{-0.21}$ & 1.14(39)  \\ 
\hline
\multicolumn{13}{c}{(C2) Single-seed Comptonization model with detectable disk} \\
\hline
Aql X$-$1     & 402053020  & {\tt diskbb+compPS[bbody]}  & $3.07\times10^{36}/1.5\times10^{-2}$  & 0.36(fixed)    &   $0.28\pm0.02$	   & $23^{+4}_{-3}$    & $0.53\pm0.02$ & $9\pm1$  & $64\pm6$ & $0.81\pm0.04$                        & $0.51\pm0.08$ & 1.15(363) \\   
              & 402053030  & {\tt diskbb+compPS[bbody]}  & $3.57\times10^{36}/1.7\times10^{-2}$  & 0.36(fixed)    &   $0.28\pm0.02$	  & $27\pm4$  & $0.52\pm0.02$ & $10\pm1$ & $51\pm5$ & $0.98^{+0.06}_{-0.05}$                       & $0.51^{+0.09}_{-0.08}$ & 1.09(363) \\
	      & 402053040  & {\tt diskbb+compPS[bbody]}  & $2.75\times10^{36}/1.2\times10^{-2}$  & 0.36(fixed)    &   $0.28^{+0.07}_{-0.05}$  & $15^{+7}_{-5}$    & $0.51\pm0.02$ & $9\pm1$  & $64\pm6$ & $0.81\pm0.04$              & $0.51\pm0.08$ & 1.16(363) \\
4U 1608$-$52  & 404044030  & {\tt diskbb+nthcomp[bbody]} & $6.47\times10^{36}/3.1\times10^{-2}$  & $1.10\pm0.04$  &   $0.55^{+0.03}_{-0.04}$  & $16\pm1$	       & $0.9\pm0.1$  &  $4\pm1$ & $22^{+78}_{-8}$  & $2.5^{+0.8}_{-1.7}$                  & $0.75^{+0.05}_{-0.06}$ & 1.24(376) \\   
              & 404044040  & {\tt diskbb+compPS[bbody]}  & $1.84\times10^{36}/8.8\times10^{-3}$  & $1.1\pm0.1$    &   $0.32^{+0.1}_{-0.06}$   & $17^{+13}_{-7}$   & $0.47^{+0.07}_{-0.04}$  & $8^{+1}_{-2}$  & $54^{+8}_{-13}$  & $1.0^{+0.3}_{-0.1}$ & $0.56^{+0.34}_{-0.19}$ & 1.20(331) \\
4U 1636$-$536 & 401050010  & {\tt diskbb+nthcomp[bbody]} & $9.71\times10^{36}/4.6\times10^{-2}$  & $0.19\pm0.01$  &   $0.55\pm0.02$	   & $14\pm1$	       & $0.86\pm0.05$ & $6\pm1$ & $20^{+5}_{-3}$ & $3.2^{+0.4}_{-0.3}$                    & $0.96\pm0.02$ & 1.15(380) \\
4U 1812$-$12  & 406008010  & {\tt diskbb+compPS[bbody]}  & $4.63\times10^{36}/2.2\times10^{-2}$  & $2.0\pm0.2$    &   $0.24^{+0.05}_{-0.03}$  & $50^{+40}_{-30}$  & $0.51\pm0.02$ & $10\pm1$ & $83^{+8}_{-9}$  & $0.80^{+0.10}_{-0.08}$               & $0.65^{+0.17}_{-0.14}$ & 1.13(331) \\ 
\hline
\multicolumn{13}{c}{(C3) Double-seed Comptonization model with common corona} \\
\hline
Aql X$-$1     & 406010010  & {\tt nthcomp[diskbb+bbody]} & $1.49\times10^{37}/7.1\times10^{-2}$  & 0.36(fixed)    &   $0.27^{+0.07}_{-0.06}$  & $40^{+24}_{-14}$  & $0.62\pm0.03$ & $11\pm1$ & $21\pm1$ & $3.60\pm0.07$           & $1.24\pm0.02$ & 1.02(330) \\  
4U 1705$-$44  & 401046010  & {\tt nthcomp[diskbb+bbody]} & $1.13\times10^{37}/5.4\times10^{-2}$  & $1.6\pm0.1$    &   0.13($<0.36$)	   & 68($>5$)	       & $0.62\pm0.01$ & $11\pm1$ & $23^{+4}_{-3}$  & $3.4^{+0.3}_{-0.2}$ & $1.24\pm0.02$ & 1.10(326) \\ 
	      & 406076010  & {\tt nthcomp[diskbb+bbody]} & $6.84\times10^{36}/3.3\times10^{-2}$  & 1.6(fixed)     &   0.13(fixed)  	   & $85\pm5$	       & $0.59\pm0.01$ & $10\pm1$ & $19^{+4}_{-2}$  & $3.7^{+0.4}_{-0.2}$ & $1.15\pm0.02$ & 1.06(314) \\  
GS 1826$-$238\tablenotemark{c} & 404007010 & {\tt dkbbfth+nthcomp[bbody]} & $1.5\times10^{37}/7\times10^{-2}$ & 0.28(fixed)  & $0.42^{+0.08}_{-0.20}$  & $>21$  & $0.63^{+0.01}_{-0.02}$ & $11.9\pm0.3$ & $>50$ & $<1.9$  & $1.21^{+0.08}_{-0.03}$ & 1.01(291) \\  
     
\hline 
\end{tabular}
\tablenotemark{a} The unabsorbed bolometric luminosity derived from the best-fit modeling. The eddington luminosity $L_{\rm edd}$ is $2.1\times10^{38}$ erg s$^{-1}$ for the canonical NS mass of 1.4$M_\odot$. \\
\tablenotemark{b} $R_{\rm in}$ is corrected for the inclination factor $\sqrt{\cos(i)}$, assuming $i=45^{\circ}$.\\
\tablenotemark{c} Results of this source are quoted from \citet{Ono16}, in which a partially Comptonized disk blackbody model {\tt dkbbfth} is applied. \\
\end{table}
\clearpage
\end{landscape}

%%%%%%%%%%%%%%%%%%%%%%%%%%%%%%%%%%%%%%%%%%
\clearpage
%========
%figure1
%========
\begin{figure}
\epsscale{1.0}
\begin{center}
\includegraphics[width=7.4cm,angle=0]{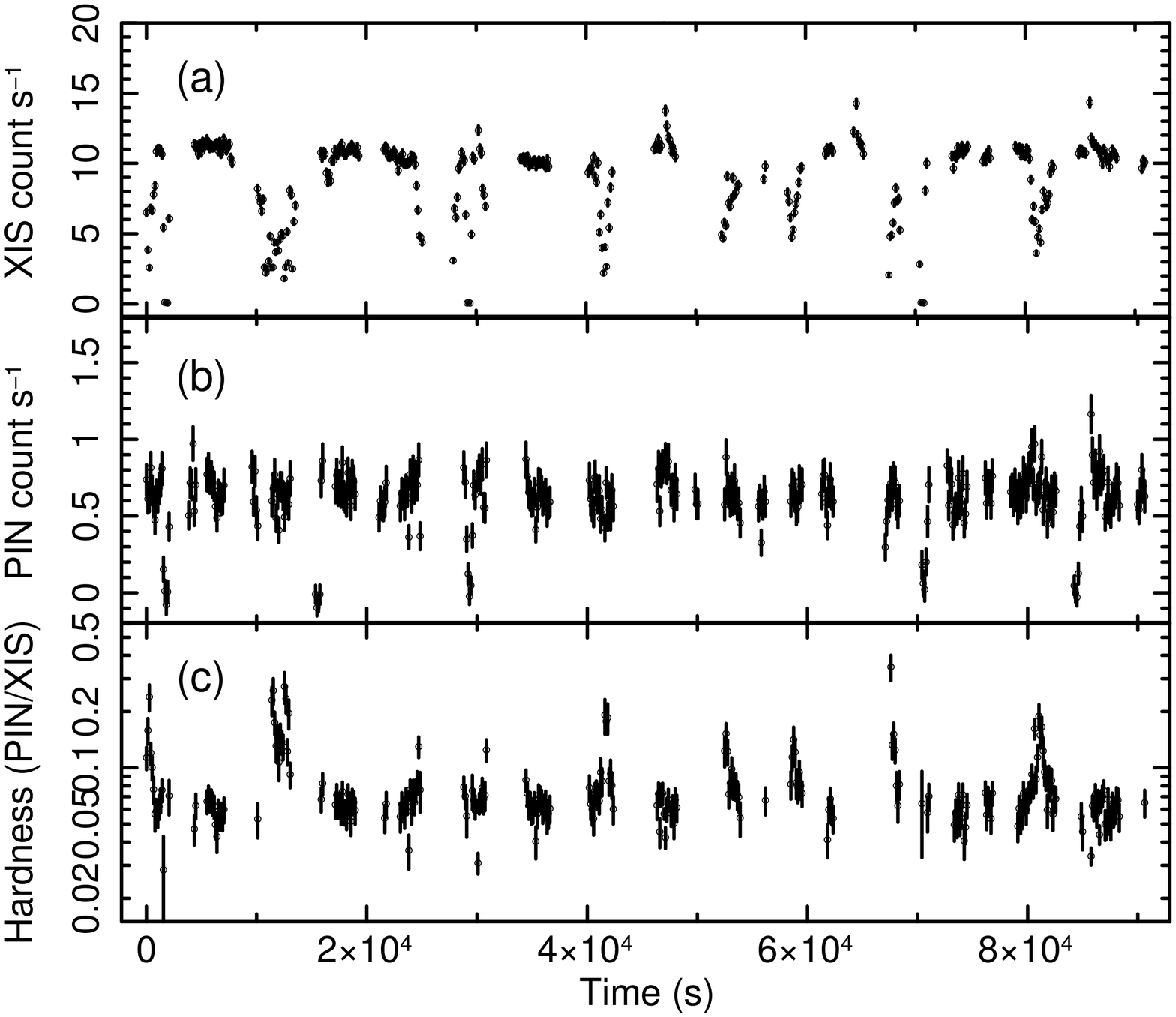}
\includegraphics[width=7.0cm,angle=0]{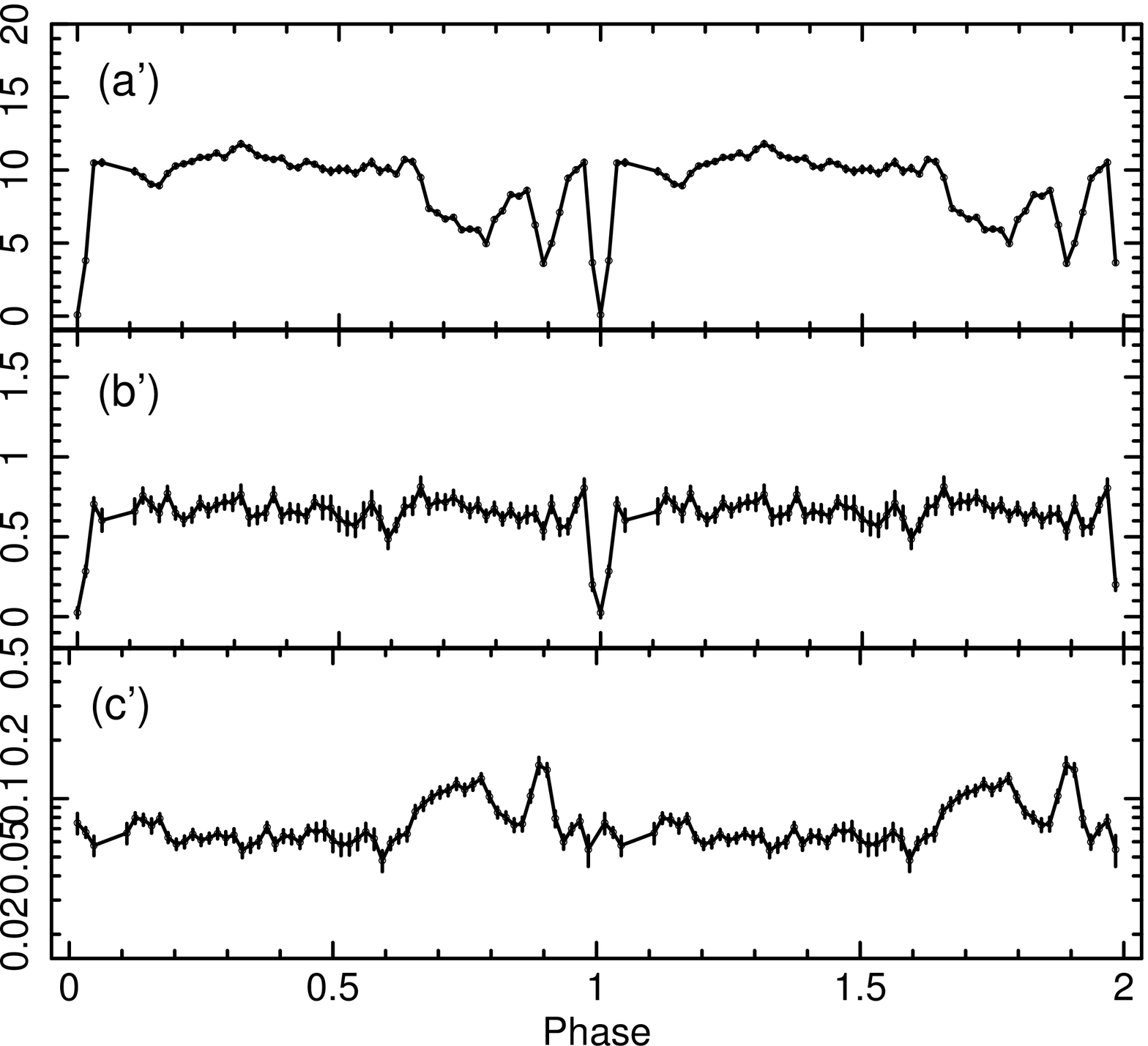}
\end{center}
\label{fig:lightcurve}
\figcaption{\suzaku~light curves and hardness ratio of EXO 0748$-$676 with a time binsize of 128 s (left panels), and those folded on $P_{\rm orb}=13766.8$ s with 64 phase bins per period after excluding the bursts (right panels), where phase 0 $(=1)$ refers to 06:12:03 of 2007 December 25. Panels (a) and (a$^{\prime}$) show the background-subtracted 0.6 $-$ 10 keV XIS light curves (averaged between XIS 0 and XIS 3). Panels (b) and (b$^{\prime}$) show NXB-subtracted 12 $-$ 55 keV HXD-PIN light curves, including the CXB contribution by $\sim$ 0.02 cts s$^{-1}$. The HXD-PIN vs. XIS hardness ratios are shown in (c) and (c$^{\prime}$).}
\end{figure}

%========
%figure2
%========
\begin{figure}
\epsscale{1.0}
\begin{center}
\includegraphics[width=11cm,angle=0]{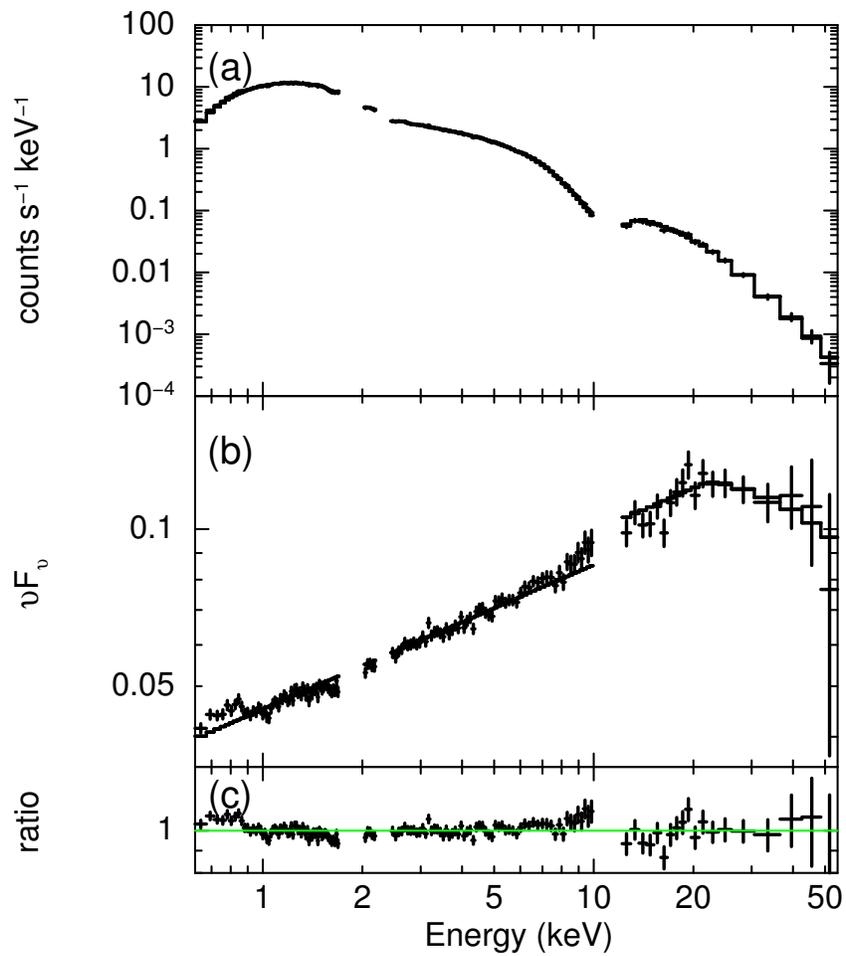}
\end{center}
\label{fig:powerlaw}
\figcaption{The persistent \suzaku~spectrum of EXO 0748$-$676 and its approximation by an empirical model {\tt highecut$\ast$powerlaw}. (a) The count spectrum and the best-fit model. (b) The deconvolved $\nu F_{\nu}$ plot and the fitted empirical model. (c) The ratio of the spectrum to the model.}
\end{figure}

%========
%figure3
%========
\begin{figure}
\epsscale{1.0}
\begin{center}
\includegraphics[width=7.7cm,angle=0]{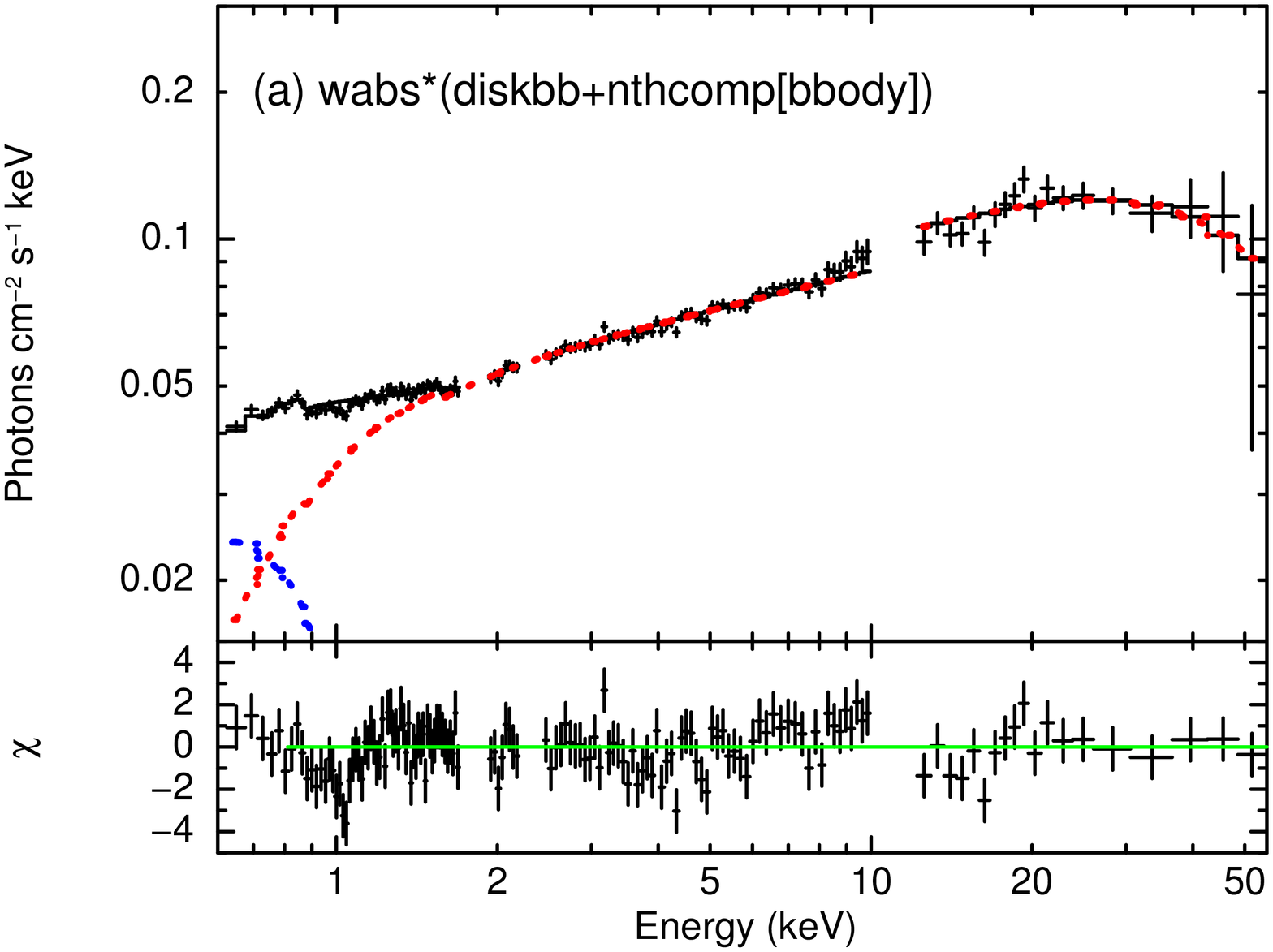}
\includegraphics[width=7.0cm,angle=0]{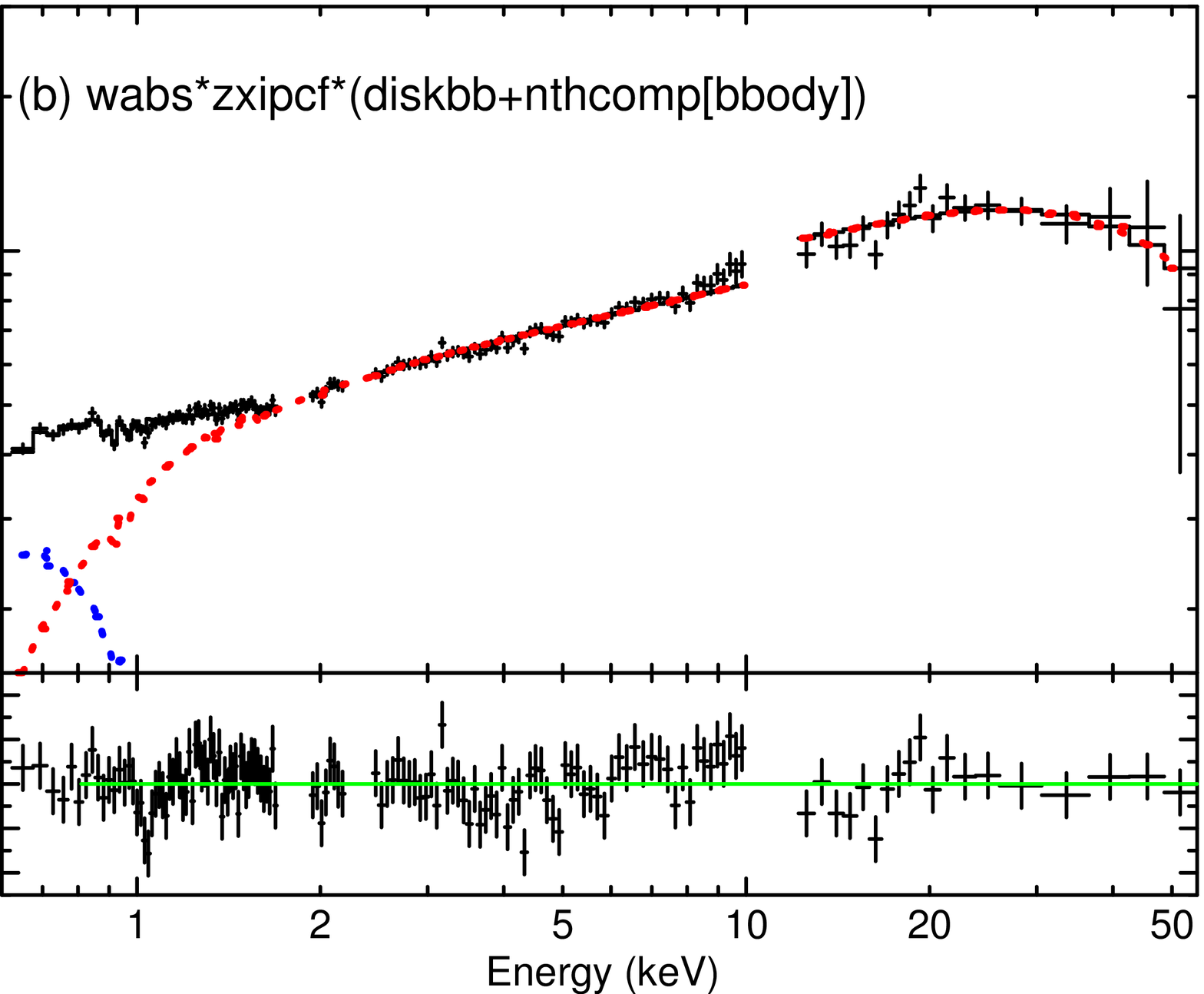}
\includegraphics[width=7.7cm,angle=0]{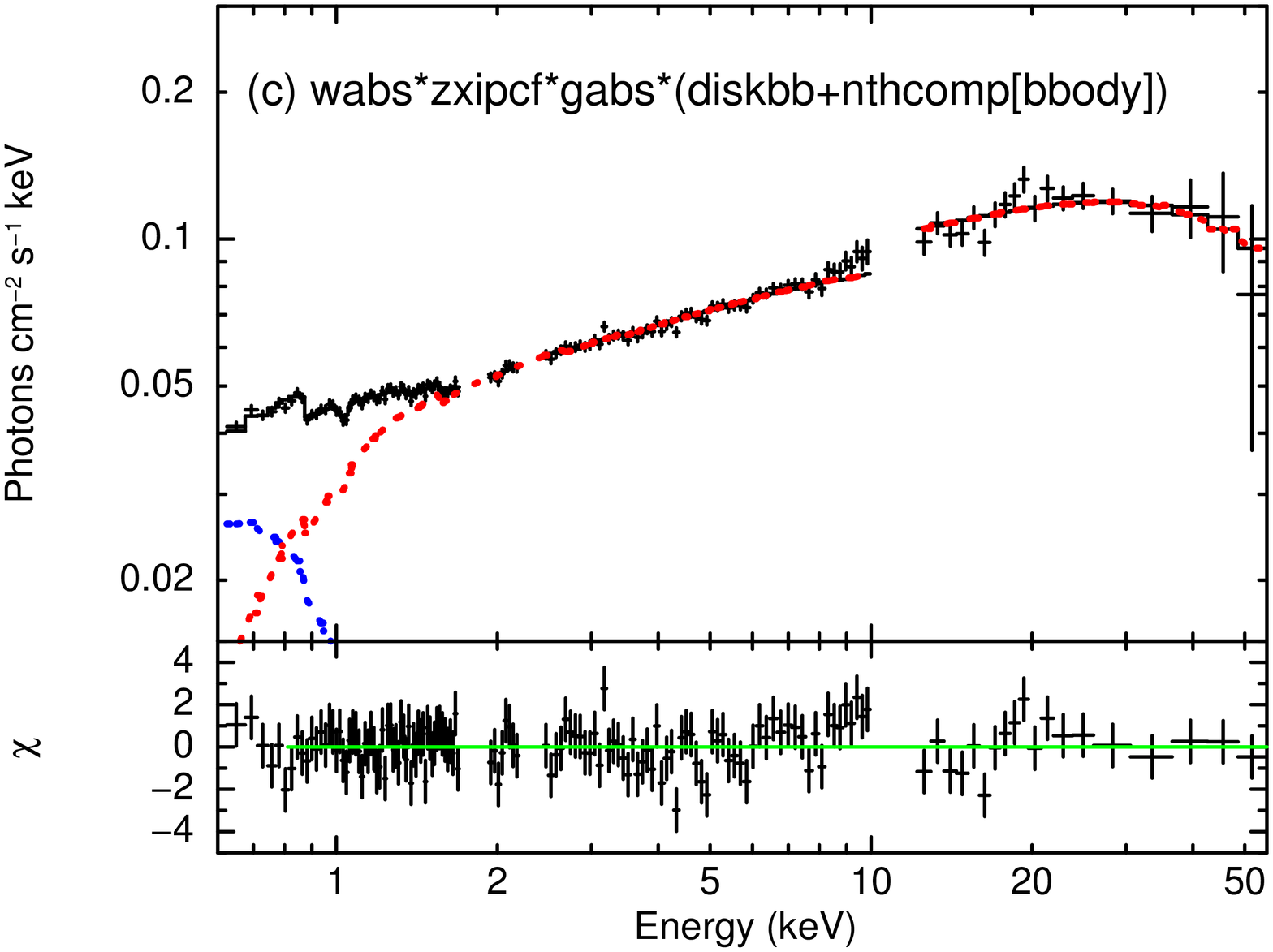}
\includegraphics[width=7.0cm,angle=0]{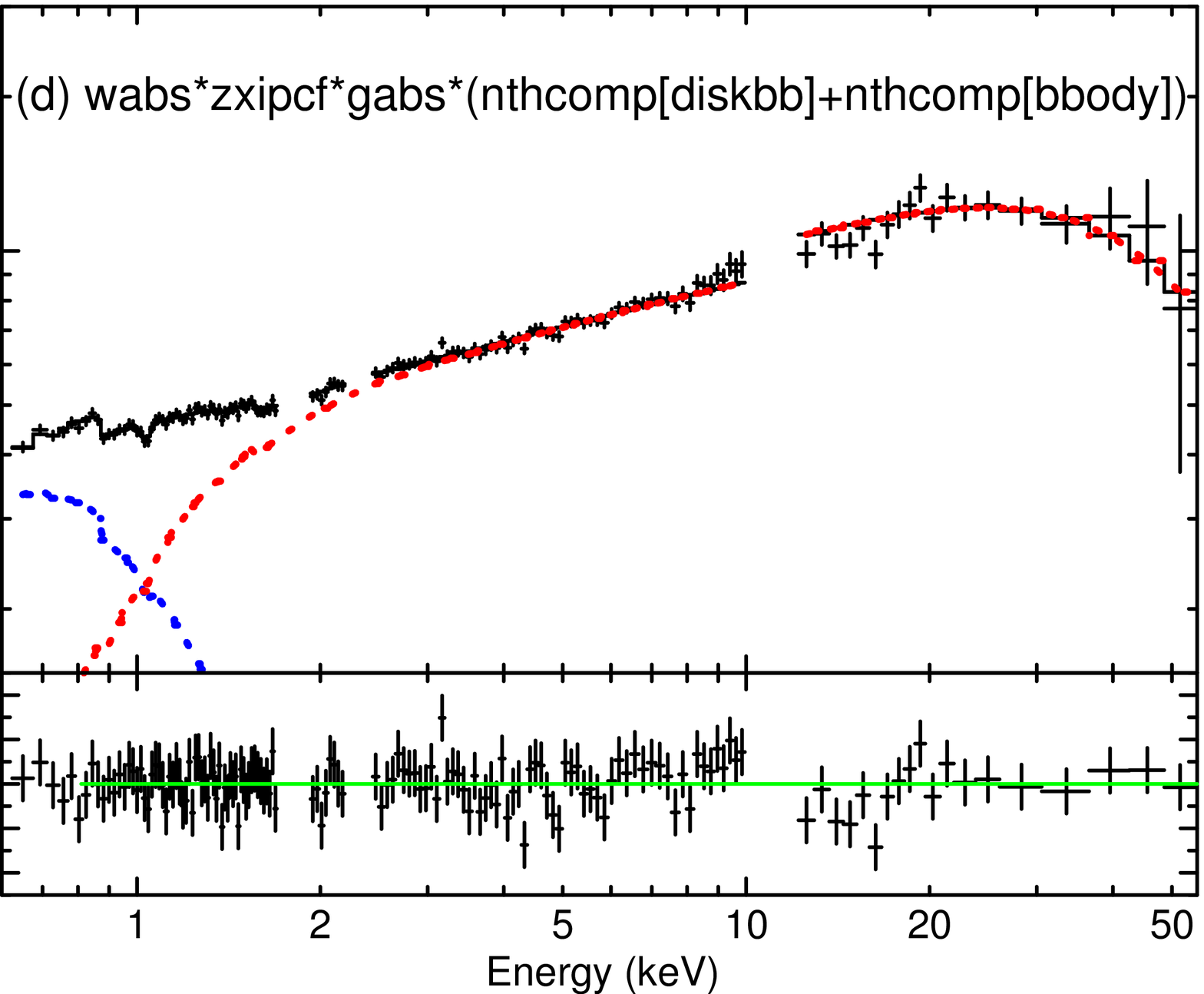}
\end{center}
\figcaption{The deconvolved $\nu F_{\nu}$ spectrum of EXO 0748$-$676 and the results of fitting with physical models. (a) Fit with a {\tt diskbb} (soft component) plus {\tt nthcomp[bbody]} (hard component) model with Galactic absorption, and its residuals. (b) Improvement of panel (a) by multiplying a partial ionized absorber {\tt zxipcf} \citep{Reeves08}. (c) Improvement of panel (b) by including a Gaussian absorption feature at $\sim$ 1 keV. (d) A double-seed Comptonization solution employing an nthcomp[diskbb] (soft component) plus nthcomp[bbody] (hard component) model. The two Comptonized components have a common electron temperature but different optical depths.}
\label{fig:spectrum}
\end{figure}

%========
%figure4
%========
\begin{figure}
\epsscale{1.0}
\begin{center}
\includegraphics[width=15cm]{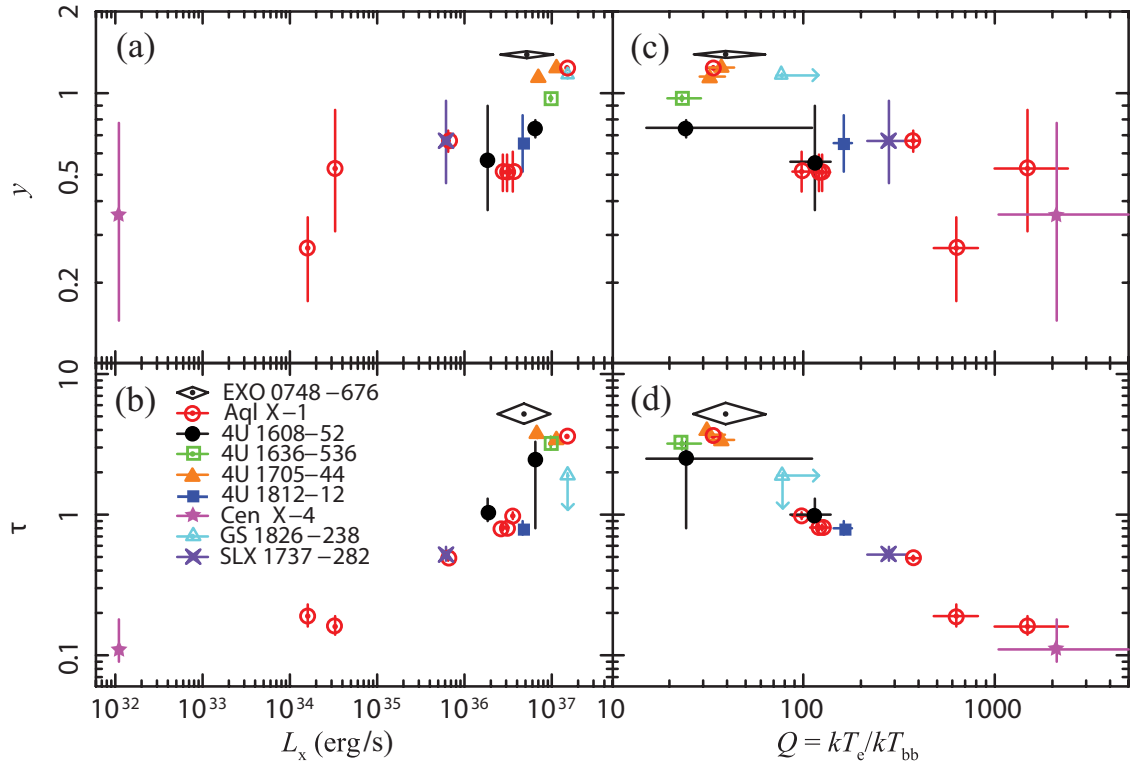}
\end{center}
\label{fig:lxQ}
\figcaption{The $L_{\rm X}$-$y$ (panel a) and $L_{\rm X}$-$\tau$ (panel b) plots for the Comptonization of the NS blackbodies in the LHS LMXBs given in table \ref{tab:sampleresult}. Panels (c) and (d) give $Q$-$y$ and $Q$-$\tau$ plots, respectively.}
\end{figure}

\end{document}